\documentstyle[12pt,psfig]{article}

\def\baselinestretch{1.2}
\def\beq{\begin{equation}}
\def\eeq{\end{equation}}

\begin{document}

\begin{titlepage}

\begin{flushright}
hep-th/0005055
\end{flushright}
\vfil\vfil

\begin{center}

{\Large {\bf Anti-Periodic Boundary Conditions in  \\Supersymmetric DLCQ }}

\vfil
{\bf S.~Pinsky and U.~Trittmann}
\\

\vfil
{\em Department of Physics \\
Ohio State University \\
Columbus, OH 43210, USA\\
}
\vfil
\end{center}

\begin{abstract}
\noindent It is of considerable
importance to have a numerical method for solving supersymmetric theories that
can support a non-zero central charge. The central
charge in supersymmetric theories is in general a boundary integral
and therefore
vanishes when one uses periodic boundary conditions. One is therefore
prevented from
studying BPS states in the standard supersymmetric formulation of
DLCQ (SDLCQ). We
present a novel formulation of SDLCQ where the fields satisfy anti-periodic
boundary conditions.
The Hamiltonian is written as the anti-commutator of two
charges, as in
SDLCQ. The anti-periodic SDLCQ we consider breaks
supersymmetry at finite resolution,
but requires no renormalization and becomes
supersymmetric in the continuum limit. In principle, this method could
be used to study
BPS states. However, we find its convergence to be disappointingly slow.
\end{abstract}

\vfil\vfil\vfil

\end{titlepage}
\renewcommand{\baselinestretch}{1.05}  

\section{Introduction}
Supersymmetric gauge theories in low dimensions
have been shown to be related to non-perturbative objects
in M/string theory \cite{witt95}, and are therefore of
particular interest nowadays. More dramatically, there is a
growing body of evidence suggesting that
gauged matrix models in $0+1$ and $1+1$ dimensions
may offer a non-perturbative formulation of string theory
\cite{bfss97,dvv97}. There is also a suggestion that large $N$ gauge
theories in various dimensions may be related to theories with gravity
\cite{mal97,ahlp99,hlpt00}.

It is therefore interesting to study directly the non-perturbative
properties of a class of supersymmetric matrix models at finite
and large $N_c$, where $N_c$ is the number of gauge colors.
Supersymmetric Discrete
Light Cone Quantization (SDLCQ)\cite{mss95} is a unique
non-perturbative numerical
approximation that is manifestly supersymmetric at each stage of the 
calculation.
In simplest terms
SDLCQ is the approximation that arises when a theory is confined to a box of
length $L$ in the spatial light-cone directions. This leads to a
discrete Fock space
basis and the supercharges are finite-dimensional matrices in this basis. This
combination of the DLCQ method and supersymmetry is well defined and
has proven to be a powerful tool which has allowed to
solve a large class of problem that have not been solved
previously \cite{ahlp99,lup99,alpp98,hak95}. It appears that
supersymmetric theories are completely well defined,
when formulated on a compact space \cite{bil99,hop96}.
All of the models that have been addressed so far have had sufficient
supersymmetry to make them
completely finite so that no renormalization was necessary. It is
important to note that
numerical approximations that do break supersymmetry will still be faced with a
renormalization problem.

In the light-cone formulation the supercharges $Q^+_\alpha$ and $Q^-_\alpha$
have several interesting and unique properties. Consider for example a pure
Yang-Mills theory in $D$ dimensions, which has a boson multiplet and a
fermion multiplet.  Since the longitudinal momentum operator is a
kinematic operator,
the supercharge
$Q^+_\alpha$ must be quadratic in the fields, while the supercharge
$Q^-_\alpha$, whose square is the Hamiltonian, in general has
both  quadratic and cubic terms in the fields. The dynamics are carried by the
cubic terms in the supercharge in the sense that a theory with only quadratic
terms in $Q^-_\alpha$ will be a non-interacting theory. The
supercharges contain an odd number of fermion fields therefore the
fermion fields
must be periodic\footnote {We will
not consider the possibility of twisted boundary conditions here, but clearly
this is an interesting direction to explore.}.

We have formulated a number of supersymmetric theories imposing 
periodic boundary
conditions on the fields and compared our results with other numerical results.
We find excellent agreement and we find the SDLCQ converges extremely fast,
much faster than any version of standard
DLCQ \cite{dlcqpapers,BPP}. Among the most
interesting results we found were a number of exactly massless
bound states in some
theories. These are states that are destroyed by one supercharge, $Q^-_\alpha$,
but not the other, $Q^+_\alpha$, since they have finite momentum. 
These massless
states persist at all values of the coupling
and it is clear that these states are BPS
states which saturate the bound $|Z|=M$ where the central charge $Z$
is zero. It would be
extremely interesting to find BPS states with non-zero masses numerically.

\section{Formulation of the Theory}
In a light-cone formulation the algebra of the supercharges with a
central charge extension takes the form
\begin{eqnarray}\label{algebra}
\{ Q^-_\alpha ,Q^-_\beta\} &=&P^- \delta_{\alpha,\beta} \\
\{ Q^+_\alpha ,Q^+_\beta\} &=&P^+ \delta_{\alpha,\beta} \nonumber  \\
\{ Q^+_\alpha ,Q^-_\beta\} &=&P_{\perp}
\gamma^{\perp}_{\alpha,\beta}+Z\delta_{\alpha,\beta},\nonumber
\end{eqnarray}
where $P^+$ is the longitudinal momentum, $P_{\perp}$ is the
transverse momentum and $P^-$ is the Hamiltonian. $Z$ is the central
charge extension and we have suppressed the spinor indices. In 
light-cone quantized
field theories one has
transverse boost invariance so one can always work in a frame where
the total transverse
momentum is zero, thus $P_{\perp}$
on all physical states can be taken to be zero.

It is well known that the central charge extension $Z$ can be written
as an integral
over the boundary of the space, and it will therefore vanish if one uses
periodic boundary
conditions. Therefore the last anti-commutator in Eq.~(\ref{algebra})
always takes the form
\begin{equation}
\{ Q^+_\alpha ,Q^-_\beta\}=0.
\end{equation}
Without a central charge extension the BPS states of a theory will
simply be massless
states. The mass of these states is protected by the BPS symmetry
and they will remain massless at all
couplings. We have seen these states in a recent SDLCQ calculation in 2+1
dimensions \cite{hhlpt99}. This theory is easily extended to an ${\cal N}{=}2$
supersymmetry,
but without a central charge there is no hope of seeing massive BPS
states. A massive BPS
state would be a very striking feature in the spectrum the theory,
since it would have a fixed mass as a function of the coupling.

We propose therefore
to extend the definition of the supercharge in order to include
anti-periodic boundary conditions. This extension breaks the
supersymmetry at finite
resolution, but at infinite resolution it will be restored.
We define momentum-shifted analogues of the standard supercharges,
$Q^-_{\pm \frac{1}{2}}$, which carry momentum $\pm \frac{\pi}{2L}$.
The Hamiltonian $P^-$ is defined by
\begin{equation}
\{ Q^-_{+\frac{1}{2},\alpha} ,Q^-_{-\frac{1}{2},\beta}\} =
2 \sqrt{2}P^- \delta_{\alpha,\beta},
\label{ham}
\end{equation}
with the normalization of Ref.~\cite{mss95}.
To test this method {\em in praxi},
we study the eigenvalue problem
\beq\label{EVP}
2P^+P^-|\psi\rangle=M^2|\psi\rangle,
\eeq
defined by the Hamiltonian, Eq.~(\ref{ham}),
within the context of the supersymmetric theory of adjoint
fermions in $1+1$ dimensions.
This is one of the simplest possible supersymmetric theories
and all the low energy bound states of this theory are well
known \cite{Kutasov93}.
The supercharge density is $q^-(x)$
\begin{equation}
q^-(x)= ig2^{7/4} Tr\left[\psi(x)\psi(x)\frac{1}{\partial_-}\psi(x)\right].
\end{equation}
where $\psi(x)$ is a real adjoint fermion field.
We expand the field into its modes which will carry half-integer momenta,
because we impose anti-periodic boundary conditions
\begin{equation}
\psi(x)_{ij} =\sqrt{{1 \over 4L}}
\sum_{n=\frac{1}{2}, \frac{3}{2}, \frac{5}{2},\ldots}
\left(B_{ij}(n)e^{-i n \pi x^-/L}
+B^\dagger_{ji}(n)e^{i n\pi x^-/L}\right).
\end{equation}
We define $Q^-_{\pm \frac{1}{2}}$ to be
\begin{eqnarray}\label{Qminus}
Q^-_{\pm \frac{1}{2}}
&=&{1\over 2L}\int_{-L}^L dx^- e^{\mp\pi x^-/2L }q^-(x)\\
&=&\frac{ig}{2^{1/4}}\sqrt{\frac{L}{\pi}}
\sum_{n_1,n_2=\frac{1}{2}, \frac{3}{2}, \frac{5}{2},\ldots}
\left(\frac{1}{n_1}-\frac{1}{n_2}-\frac{1}{n_1+n_2 \mp
\frac{1}{2}}\right)\times\nonumber\\
&\times&\left(B^\dagger_{ik}(n_1)B^\dagger_{kj}(n_2)
B_{ij}(n_1+n_2 \mp \frac{1}{2})+
B^\dagger_{ij}(n_1+n_2 \pm \frac{1}{2})B_{ik}(n_1)B_{kj}(n_2)\right).
\nonumber
\end{eqnarray}
Since we are using a real representation for the fermions it is easy
to verify that the Fock space matrices $Q^-_{\pm\frac{1}{2}}$ satisfy,
\begin{equation}
Q^-_{+\frac{1}{2}}=\left[Q^-_{-\frac{1}{2}}\right]^t.
\label{trans}
\end{equation}

If all the wavefunctions of the physical states vanish sufficiently
fast for small momentum, then for large resolution one would expect
that the half unit of momentum
which the supercharges carry is negligible compared to momentum the
partons carry. Thus,
$Q^-_{\pm \frac{1}{2}} \rightarrow Q^-$ in the continuum limit
$K\rightarrow \infty$.
However, since we have broken the supersymmetry one might face a
mass renormalization which is generally required
in these two-dimensional theories. In spite of this potential
difficulty we proceed to calculate the spectrum of this theory, and
will find that the theory indeed need not be renormalized.

\section{Numerical Results}

In order to obtain the mass squared eigenvalues $M^2$ of the theory,
we have to solve the eigenvalue problem, Eq.~(\ref{EVP}).
We use the standard large $N_c$ discrete Fock basis. Its states are of
the form
\[
| p_1,p_2,\ldots, p_n\rangle=\frac{1}{{N_c}^{n/2}\sqrt{s}}
Tr\left[B^{\dagger}(p_1)
B^{\dagger}(p_2)\cdots B^{\dagger}(p_n)\right]|0\rangle,\\
\]
where the symmetry factor $s$ counts the number of times the set of momenta
$(p_1,\ldots,p_n)$ is mapped onto itself under cyclic permutations.
When using anti-periodic boundary conditions, the fermion bound states
will lie in the sector of half-integer harmonic resolution $K$ and boson
bound states will have integer $K$.
The action of the operators, Eq.~(\ref{Qminus}), can be readily evaluated
\begin{eqnarray*}
&&Q^-_{\pm \frac{1}{2}}| p_1,p_2,\ldots, p_n\rangle=
\frac{ig\sqrt{L}}{2^{7/4}}\sqrt{\frac{N_c}{\pi}}\sum_{i=1}^{n}(-)^{i+1}
\\
&&\quad\quad
\left\{(-)^{n(i+1)}\sum_{k=\frac{1}{2}}^{p_i\pm 1}
\left(\frac{1}{k}+\frac{1}{p_i-k\pm \frac{1}{2}}-\frac{1}{p_i}\right)
| k,p_i-k\pm \frac{1}{2},L_n^{(i-1)}(p_i)\rangle\right.\\
&&\quad\quad
\left.
+(-)^{ni}\left(\frac{1}{p_{i-1}}+\frac{1}{p_i}
-\frac{1}{p_i+p_{i-1}\pm \frac{1}{2}}\right)
|p_i+p_{i-1}\pm \frac{1}{2},L_n^{(i-1)}(p_i,p_{i-1})\rangle\right\},\\
\end{eqnarray*}
where $L^{(j)}_n$ is a permutation of $n-k$ momenta
\[
L^{(j)}_n(p_{i_1},\ldots,p_{i_k})=
\{p_{1+j},p_{2+j},\ldots,p_{i_1-1+j},p_{i_1+1+j},\ldots,
p_{i_k-1+j},p_{i_k+1+j},\ldots p_{n+j}\}.
\]
This compact form allows for an easy  computer implementation.
Since the supercharges change fermions to bosons and vice versa, they change
the resolution $K$ by $\pm \frac{1}{2}$.
Thus  in this discrete representation the supercharge matrix,
\[
\langle K|Q^-_{\pm \frac{1}{2}}|K\mp \frac{1}{2}\rangle,
\]
will in general not be a square matrix.
The Hamiltonian, however, is constructed as the anti-commutator of the
of the two momentum-shifted supercharge matrices, Eq.~(\ref{ham}).
 From Eq.~(\ref{trans}) it follows then that the Hamiltonian
is a hermitian matrix. In this matrix representation the Hamiltonian
at resolution $K$ will receive contributions from intermediate states
with resolution $K\pm \frac{1}{2}$.
The supercharge matrices in a $(fermion,boson)^t$ basis have the structure
\begin{eqnarray}
Q^-_{\pm\frac{1}{2}}=\left(
\begin{array}{cc}
\mbox{\bf 0} & A_{\pm\frac{1}{2}}\\
B_{\pm\frac{1}{2}} & \mbox{\bf 0}
\end{array}
\right)
\end{eqnarray}
The Hamiltonian for the fermions is therefore,
\beq\label{Pferm}
P^-_{ferm}=A_{+\frac{1}{2}}B_{-\frac{1}{2}}+A_{-\frac{1}{2}}B_{+\frac{1}{2}},
\eeq
and the Hamiltonian for the bosons is,
\beq\label{Pbose}
P^-_{bose}=B_{+\frac{1}{2}}A_{-\frac{1}{2}}+B_{-\frac{1}{2}}A_{+\frac{1}{2}}.
\eeq
These two matrices are subsequently diagonalized to yield the fermionic
the bosonic masses. We note that these matrices have in general
different dimensions.

%
\begin{figure}
\centerline{
\psfig{file=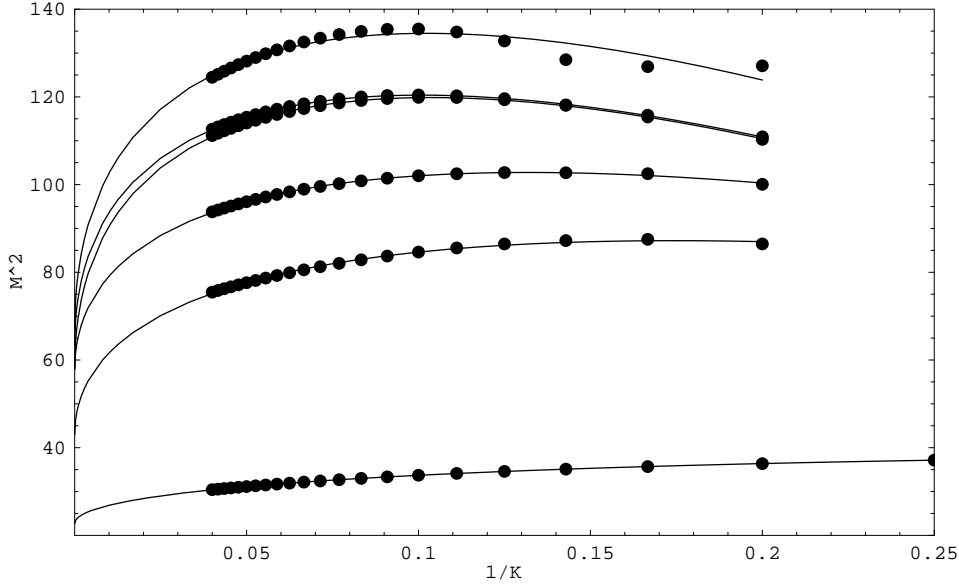,width=5in}}
\caption{The six lowest masses in units of $g^2N_c/\pi$ as a function
of the inverse of the harmonic resolution ${1\over K}$.
All data were fitted to a function of the form $M^2 + a\sqrt{1\over K} +
b {1\over K}$.
\label{fig1}}
\end{figure}
In a first step, we calculated 
the fermionic and bosonic spectrum up to harmonic resolution $K{=}10$
by solving the eigenvalue problem, Eq.~(\ref{EVP}).
We then focused on the six lowest eigenvalues whose eigenfunctions
for resolutions up to $K{=}10$ are solely built out of
states with at most four and five particles
for the bosonic and fermionic spectrum, respectively.
A truncation to four particles allowed us to go to resolution
$K{=}25$ in the bosonic sector with relative ease.
In the fermionic sector of the spectrum, the larger
dimension of the fermionic matrix, Eq.~(\ref{Pferm}), prevented us 
from going quite as high.

The resulting bound state masses in units of
$g^2N_c/\pi$ are shown in Fig.~\ref{fig1}.
We fitted all six curves to the function
\[
M^2 + a\sqrt{1\over K} +b {1\over K}.
\]
We find the continuum masses to be
\beq
M^2_{aSDLCQ}=22.7, 43.0, 57.7, 58.0, 63.4, 67.0,
\eeq
compared to the values
\beq
M^2_{SDLCQ}=23.8, 47.9, 62.1, 62.6, 63.8, 64.7,
\eeq
of ordinary SDLCQ at resolution $K=8$.
While the agreement is far from perfect, it is clear that for this 
model the anti-periodic SDLCQ
method works,  in the sense that it correctly
reproduces the spectrum of the theory.  It is fair to say, however, that the
exact continuum values are quite sensitive to the fitting function.

\begin{figure}
\centerline{\psfig{file=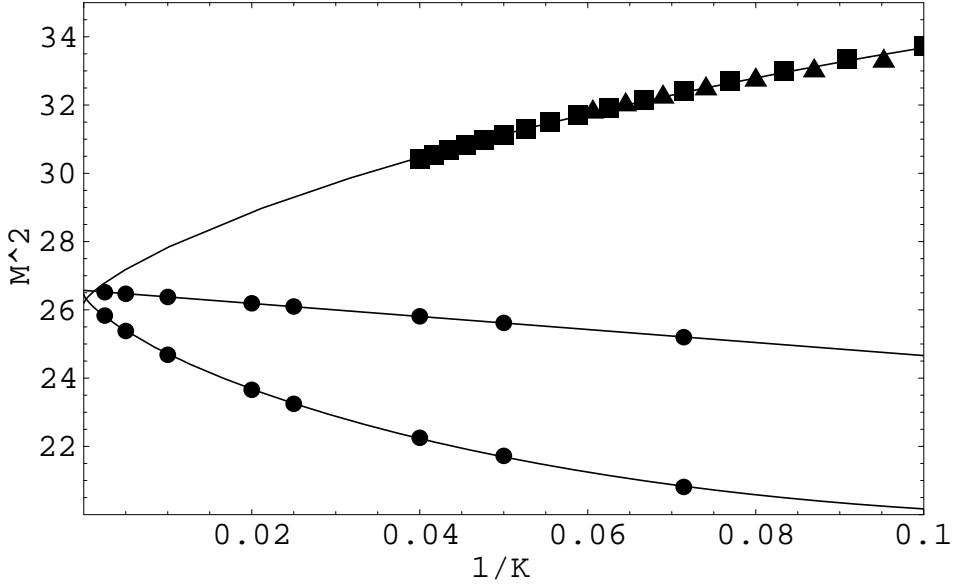,width=5in}}
\caption{
All the curves correspond to the lowest mass state (in units of $g^2
N_c/\pi$) of the
theory. The boxes and triangles represent the lowest
anti-symmetric SDLCQ boson and fermion eigenstates, respectively.
The middle curve stems from a calculation using ordinary SDLCQ and
the fermions and bosons are
exactly degenerate. The lower curve is the standard DLCQ result for the lowest
mass boson. The
upper and lower curves are both fitted to $M^2 + a' K^{-3/4} + b' K^{-3/2}$.}
\label{fig2}
\end{figure}

In Ref.~\cite{alp99} we compared the
lowest eigenvalue as calculated in SDLCQ and standard DLCQ as a function of the
resolution $K$. We found that the convergence of standard DLCQ is slow
and clearly nonlinear
as a function of $1/K$. Unfortunately, we see from Fig.~\ref{fig2}
that the convergence for
anti-periodic SDLCQ is, roughly speaking, about as bad as the standard
Hamiltonian DLCQ method\footnote{For comparison, we 
used the same fitting function 
that was used in the DLCQ calculations 
of Ref.~\cite{alp99} for our data.
}.
In Fig.~\ref{fig2} we show the lowest fermion bound state mass
and the lowest boson bound state mass as a function of $1/K$ for the
anti-periodic SDLCQ
approximation. While the fermions and bosons appear at different
resolutions, they can be fitted by
the same curve for large enough $K$.
Thus the supersymmetry of the spectrum appears to
be recovered by the approximation already at low harmonic resolution.
Unfortunately, although the potential perils of supersymmetry breaking seem to
be absent already at $K=10$ for the lowest eigenstates, the convergence
remains unpleasantly slow.
In principle, however, the method appears to be working, and we anticipate
room for numerical improvements.

\section{Conclusions}

In the present note we introduced a novel
version of SDLCQ that allows for the use of anti-periodic boundary conditions
for all fields. Consequently, the formalism can support for the first time
boundary integral contributions. We provided evidence that
the approximation converges to the standard supersymmetric results.
The new approach inherits the very advantageous feature of
absence of mass renormalization from SDLCQ. Unfortunately however, the
approximation does not enjoy the same rapid convergence as SDLCQ. The lack
of rapid convergence is a severe problem, since supersymmetric theories with
non-trivial central charges (${\cal N}\geq 2$) have many species of particles
and therefore a very
large Fock space. Thus going to sufficiently high resolution to
obtain accurate results for the spectrum will be quite difficult.
Of course, the standard DLCQ approach can be used with anti-periodic
boundary conditions. But,
as we saw in Fig.~\ref{fig2}, its convergence is no better than that of
the anti-periodic SDLCQ approximation.
Moreover, the standard DLCQ approach will certainly have
renormalization problems in higher dimensions.

\section*{Acknowledgments}

This work was supported in part by the US Department of Energy. The 
authors would like to thank
Oleg Lunin for many helpful discussions.

\end{document}